# Protein and DNA sequence determinants of thermophilic adaptation

Konstantin B. Zeldovich, Igor N. Berezovsky, Eugene I. Shakhnovich

Department of Chemistry and Chemical Biology, Harvard University, 12 Oxford St, Cambridge, MA 02138

E-mail: eugene@belok.harvard.edu

## Abstract

**Background.** There have been considerable attempts in the past to relate phenotypic trait - habitat temperature of organisms - to their genotypes, most importantly compositions of their genomes and proteomes. However, despite accumulation of anecdotal evidence, an exact and conclusive relationship between the former and the latter have been elusive.

**Methodology/Main Findings.** We present an exhaustive study of the relationship between amino acid composition of proteomes, nucleotide composition of DNA, and optimal growth temperature of prokaryotes. Based on 204 complete proteomes of archaea and bacteria spanning the temperature range from -10C to +110C, we performed an exhaustive enumeration of all possible sets of amino acids and found a set of amino acids whose total fraction in a proteome is correlated, to a remarkable extent, with the optimal growth temperature. The universal set is Ile, Val, Tyr, Trp, Arg, Glu, Leu (IVYWREL), and the correlation coefficient is as high as 0.93. We also found that the G+C content in 204 complete genomes does not exhibit a significant correlation with optimal growth temperature (R=-0.10). On the other hand, the fraction of A+G in coding DNA is correlated with temperature, to a considerable extent, due to codon patterns of IVYWREL amino acids. Further, we found strong and independent correlation between OGT and frequency with which pairs of A and G nucleotides appear as nearest neighbors in genome sequences. This adaptation is achieved via codon bias.

**Significance** These findings present a direct link between principles of proteins structure and stability and evolutionary mechanisms of thermophylic adaptation. On the nucleotide level, the analysis provides an example of how Nature utilizes codon bias for evolutionary adaptation to extreme conditions. Together these results provide a complete picture of how compositions of proteomes and genomes in prokaryotes adjust to the extreme conditions of the environment.

**Synopsis**

Prokaryotes living at extreme environmental temperatures exhibit pronounced signatures in the amino acid composition of their proteins and nucleotide compositions of their genomes reflective of adaptation to their thermal environments. However, despite significant efforts, the definitive answer of what are the genomic and proteomic compositional determinants of Optimal Growth Temperature of prokaryotic organisms remained elusive. Here the authors performed a comprehensive analysis of amino acid and nucleotide compositional signatures of thermophylic adaptation by exhaustively evaluating all combinations of amino acids and nucleotides as possible determinants of Optimal Growth Temperature for all prokaryotic organisms with fully sequences genomes.. The authors discovered that total concentration of seven amino acids in proteomes – IVYWREL – serves as a universal proteomic predictor of Optimal Growth Temperature in prokaryotes. Resolving the old-standing controversy the authors determined that the variation in nucleotide composition (increase of purine load, or A+G content with temperature) is largely a consequence of thermal adaptation of proteins. However, the frequency with which A and G nucleotides appear as nearest neighbors in genome sequences is strongly and independently correlated with Optimal Growth Temperature. as a result of codon bias in corresponding genomes. Together these results provide a complete picture of proteomic and genomic determinants of thermophilic adaptation.

**Introduction**

As proteins and nucleic acids must remain in their native conformations at physiologically relevant temperatures, thermal adaptation requires adjustment of interactions within these biopolymers. Given the limited alphabet of amino acid residues, an apparent way to control protein stability is to properly choose the fractions of different residue types and then to arrange them in sequences that fold into and stable in unique native structures (1) (2) (3) (4) at physiological conditions of a given organism. Various mechanisms of thermostability were discussed in the literature and many authors pointed out to changes in amino acid composition as one of the clearest manifestation of thermal adaptation (5), (4) (2) (6) (7) (8) .

Indeed, it is well known that enhanced thermostability is reflected in specific trends in amino acid composition (9) (10) (11). The most pronounced ones are in elevated fraction of charged residues (12) (13) (14) (15, 16), and/or increased amount of hydrophobic residues in (hyper)thermophilic organisms as compared to mesophilic ones. An early attempt of a systematic search for amino acids that are most significant for protein thermostability was made by Ponnuswamy et al.(17), who considered a set of 30 proteins and about 65000 combinations of different amino acids to find the amino acid sets serving as best predictors of denaturation temperature.

Furthermore, a number of authors explored possible relationship between an Optimal Growth Temperature (OGT) of organisms and nucleotide content of their genomes (16, 18, 19). An increase in purine (A+G) load of  bacterial genomes of some thermophiles was noted recently (20) as a possible primary adaptation mechanism.

However, despite the significant efforts, a clear and comprehensive picture of genomic and proteomic signatures of thermal adaptation remains elusive. First, it is not fully established which compositional biases in genomes and proteomes represent the most definitive signatures of thermophilic adaptation. On the genomic level is it G+C as widely believed by many after classical experimental study by Marmur and Doty (21) or alternative signatures such as purine loading index as suggested by others (19)? For proteomic compositions is it excess of charge residues or hydrophobic or both and which amino acids specifically are most sensitive to thermal adaptation? On a more fundamental level the key question is which factor – amino acid or nucleotide composition - is primary in thermal adaptation and which one is derivative? This issue is most vivid in prokaryotes whose genomes consist mostly of coding DNA, leading to a well-defined relationship between nucleotide and amino acid compositions. However this relationship may be not rigid due to degeneracy of genetic code. In principle amino acid and nucleotide compositions can adapt, to a certain degree, independently due to possible application of codon bias.

Definite answers to these questions can be obtained only via a comprehensive study that considers in a systematic way, all known prokaryotic genomes and all possible composition factors. Here, we carry out a comprehensive investigation of the relationship between the optimal growth temperature (OGT, $T_{opt}$) of prokaryotes and the compositions of their complete proteomes and compositions and pairwise nearest-neighbor correlations in their genomes. First, our goal is to find the sets of amino acids and nucleotides whose total content in a proteome (genome) serves as the best predictor of the OGT of an organism. We perform an exhaustive enumeration of all possible combinations of amino acid residue types, without making any *a priori* assumptions of the relevance of a particular combination for thermostability. Our analysis is based on 204 complete genomes of bacteria and archaea thriving under temperatures from -10°C to 110°C, comprising psychro-, meso-, thermo- and hyperthermophilic organisms (Table S1 in Supporting Information). It reveals a particular combination of amino acids whose cumulative concentration in proteomes is remarkably well correlated with OGT. This result holds mainly for globular proteins comprising most parts of prokaryotic proteomes. For comparison, we apply this approach also to membrane proteins, having α-helical bundles (22) and β-barrels (23) structures, and show that the amino acids predictors of OGT found for proteomes of globular proteins does not work for membrane proteins. This finding clearly indicates that mechanisms of thermal stabilization of membrane proteins are different from those of globular proteins.

Next we turn to genomes and carry out the same comprehensive analysis to determine compositional genomic determinants of OGT. We find that purine load index, i.e. concentration of A+G exhibits highest correlation with OGT, consistent with an earlier observation made on several individual genomes (20). Having found both proteomic and genomic compositional characteristics that correlate with OGT we then turn to the key question of whether genomic and proteomic determinants are independent ones or one is a derivative of the other. By running a set or reshuffling controls as described below we show that primary factor is adaptation on the level of amino acid composition and the variation in DNA composition is largely a derivative of amino acid adaptation. While the nucleotide composition biases appear to be largely due to adaptation at the level of amino acid compositions, an additional DNA sequence adaptation that is independent of amino acid composition adaptation is still possible – at the level of higher order correlation in nucleotide sequences. A possible mechanism for that in prokaryotes is through codon bias. Indeed we find a clear evidence for independent adaptation in DNA at the level of nearest-neighbor correlation of nucleotides (possibly due to strengthening of stacking interactions). We show that this adaptation of DNA sequences does indeed occur via codon bias. Finally we discuss how observed patterns of change in aminoacid compositions in response to extreme conditions of the environment are related to physical principles that govern stability of globular proteins.

**Methods**

The complete genomes (Table S1) were downloaded from NCBI Genome database and the optimal growth temperatures were collected from original publications, and from the web sites of the American Type Culture Collection, http://www.atcc.org, the German Collection of Microorganisms and Cell Cultures, http://www.dsmz.de, and PGTdb database (24), http://pgtdb.csie.ncu.edu.tw/. The distribution of OGTs of prokaryotes with completely sequenced genomes shows that 75 species out of 204 have an OGT of 37°C (mostly human and animal pathogens), 27 species have an OGT of 30°C, and 19 species grow best at 26°C. Therefore, 121 genomes, or 59% of the data, correspond to just three values of OGT. To overcome this obvious bias in the data, we averaged the amino acid compositions of the organisms with OGTs of 26, 30, and 37°C respectively. After this procedure, our data set consists of 86 proteomes, 83 real proteomes and 3 averaged proteomes at 26, 30, and 37°C. To validate the averaging of the overrepresented amino acid compositions we repeated the complete analysis for the 204 genomes, and did not find any major differences, see Supporting Information.

We represent the different sets of amino acids by vectors $\{a_i\}$ of 20 binary digits, where each digit $a_i$ takes the value of 1 if the amino acid of type $i$ is present in the set, and 0 otherwise. For the set to be nontrivial, at least one of the components of $a_i$ must be 1 and at least another one must be zero. There are $2^{20}-2=1,048,574$ such vectors, enumerating all possible subsets of 20 amino acids. Let $f_i^{(j)}$ be the fraction of amino acid of type $i$ in proteome $j$. Then, for each of the 86 proteomes and each vector $\{a_i\}$ we calculate the total fraction $F^{(j)}$ of the amino acids from a particular subset,

$$F^{(j)} = \sum_{i=1}^{20} a_i f_i^{(j)}$$

and perform the linear regression between the values of $F^{(j)}$ and organism living temperatures $T_{opt}^{(j)}$. The higher the correlation coefficient $R$ of this regression, the more important is a given subset of amino acids as a predictor of OGT. Since for each proteome $\sum_{i=1}^{20} f_i = 1$, there are in fact only $2^{19}-1$ linearly independent combinations of fractions of amino acids; if the total fraction of $N$ amino acids increases with OGT and the correlation coefficient is $R$, the total fraction of the complementary set of 20-$N$ amino acids is decreasing with OGT with the correlation coefficient $-R$. Therefore, the predictive powers of the complementary subsets of amino acids are equal. To resolve this ambiguity, we focus on the amino acid combinations which are positively correlated with OGT.

For 86 proteomes, the best predictor consists of amino acids IVYWREL, correlation coefficient $R$=0.930, accuracy of prediction 8.9°C (see Results).

We also performed an exhaustive enumeration of $3^{20}$-1 combinations of amino acids in a ternary model, where the coefficients $a_i$ can take the values of -1, 0, and 1. We expected that this method could be advantageous, as it allows to enrich the predictor combination by the amino acids whose fraction is decreasing with OGT (such amino acids would have a weight of -1). The best predictor combination in the ternary model consists of amino acids VYWP with weight +1 and CFAGTSNQDH with the weight -1, correlation coefficient $R$=0.948, accuracy of prediction of 8.1°C for 86 proteomes. While this result is generally consistent with the IVYWREL predictor, the ternary model does not offer a significant increase of accuracy of the prediction of OGT as compared to a much simpler binary model. It is also possible to perform a full linear regression between the amino acid fractions and OGT, effectively allowing for non-integer values of $\{a_i\}$. This procedure involves a great risk of overfitting, and results only in a marginal improvement of accuracy of OGT prediction as compared to both binary and ternary models.

TMHMM prediction server (22) at http://www.cbs.dtu.dk/services/TMHMM/ has been used to identify the sequences of alpha-helical membrane proteins in the 83 genomes. The number of proteins predicted in a genome did not correlate with OGT, so the prediction algorithm is sufficiently robust with respect to the changes of average amino acid composition. The PROFTmb server (23) at http://cubic.bioc.columbia.edu/services/proftmb/ has been used to identify the transmembrane beta-barrels; proteins with 8 to 22 transmembrane beta-strands have been selected for further study.

We applied the same technique to enumerate all possible combinations of nucleotides in the coding DNA of prokaryotes and find the correlation coefficients between the total fractions of sets of nucleotides and OGT. To quantify the pairwise nearest-neighbor correlations in DNA, for each of the 16 possible pairs of nucleotide types $i$ and $j$ and for each of the 83 genomes we calculated the correlation function $c_{ij} = \frac{Nn_{ij}}{n_i n_j} - 1$, where $N$ is the total number of nucleotides in the DNA sequence, $n_i$, $n_j$ are the numbers of nucleotides of types $i$ and $j$, and $n_{ij}$ is the number of pairs where nucleotide $j$ immediately follows nucleotide $i$ in the coding strand. In random DNA without sequence correlations, $c_{ij}=0$ for any $i,j$ and for any nucleotide composition of the sequence.

Reshuffling of DNA and protein sequences has been performed by swapping two randomly chosen letters in the sequence, and repeating this procedure $2N$ times, where $N$ is the length of the sequence.

Sequences of 10 protein folds (according to SCOP description (25)) with less than 50% of sequence identity were extracted from PDB database. Given a fold, representatives of its sequences were found in 83 complete proteomes by using blastp program with BLOSUM62 substitution matrix, e-value equals to 0.005. For every analyzed protein fold, representatives of its sequences were detected in organisms with OGTs covering whole interval +10 to 110ºC. Extracted proteomic sequences were used in calculations of thermostability predictor if the length of the alignment was more than 70 % of the size of fold's sequence.

## Results

**Sequence determinants of thermal adaptation of soluble proteins.**

We computed the $2^{20}-2=1,048,574$ values of the correlation coefficient $R$ of the dependence between OGT $T_{opt}^{(j)}$ and the fractions $F^{(j)}$ for all possible sets of amino acids $\{a_i\}$, and found the set yielding the highest $R$. For the 86 proteomes with different OGT (see Methods) studied, the best set of amino acids is Ile, Val, Tyr, Trp, Arg, Glu, Leu (IVYWREL), giving the correlation coefficient between $F_{IVYWREL}$ of a proteome and optimal growth temperature of the organism $R=0.930$ (Figure 1). The quantitative relationship between the optimal growth temperature $T_{opt}$ (in degree Celsius) and fraction $F$ of IVYWREL amino acids reads

$$T_{opt} = 937\,F - 335. \qquad (1)$$

The root-mean-square deviation between the actual and predicted OGTs for 86 genomes is 8.9°C. The same analysis performed on the complete set of 204 genomes yields KVYWREP as the best OGT predictor, with the correlation coefficient $R=0.84$ and root-mean-square error of 9.2°C (Supplementary Figure S1). The ten best amino acid combinations for 86 genomes and the corresponding correlation coefficients are presented in Supplementary Table S2. Supplementary Table S5 presents all of the one- and two-letter amino acid combinations and their correlation coefficients $R$.

### Statistical tests and controls

We performed several tests to substantiate our procedure and to prove that the IVYWREL set is not a numerical aberration in a fit of noisy data, but a robust predictor of habitat temperature of prokaryotes.

First, as a control that the data are not overfitted, we randomly reshuffled the values of $T_{opt}^{(j)}$ between organisms, and, using the same exhaustive enumeration of all amino acid combinations, found the best composition-temperature correlation in the resulting artificially randomized proteome-temperature combinations. The distribution of the maximum correlation coefficient $R_{max}$ obtained in 1000 such reshufflings follows a Gaussian shape centered around $<R_{max}>$=0.344, $\sigma$=0.060. Therefore, the probability to find the observed correlation of $R$=0.93 as a result of overfitting in a random combination of proteomes and temperatures is less than $p$=10$^{-20}$, which proves an extremely high statistical significance of both the IVYWREL set and its correlation coefficient $R$=0.93.

Supplementary Figure S2 shows another random control - histogram of the correlation coefficient $R$ of the $F(T_{opt})$ dependence for all possible combinations of 7 amino acids (black curve) and for all variations of the IVYWREL set with one substitution of the amino acid (red curve). The correlation coefficient for the IVYWREL-like sets is significantly higher than that for a combination of seven random amino acids.

To demonstrate the robustness and stability of the IVYWREL predictor set, in Figure 2 we plot the distribution of correlation coefficients of the $F(T_{opt})$ dependence for different variations of the IVYWREL set. Figure 2 shows that the IVYWREL set is very tolerant to addition or removal of one amino acid, with the median value of $R$ in excess of 0.8. Addition or deletion of two letters, or a substitution of one letter have a more significant effect on the accuracy of OGT prediction, with median $R$ on the order of 0.75. Table 1 shows the examples of additions, deletions and substitutions in the IVYWREL set having the strongest or weakest effect on the correlation coefficient $R$.

To check the power of our method in the prediction of optimal growth temperature of prokaryotes, we randomly split the set of 86 proteomes into training sets of 43 and a test set of 43 species. We used the training set to determine the parameters of the best linear relationship between the fractions of all $2^{20}$-1 amino acid sets and OGT, and employed this regression to predict OGTs of organisms in the test set. The accuracy of this prediction can be characterized by the root-mean-square difference $\sigma_{\Delta T}$ between predicted and actual OGT of the 43 organisms in the test set,

$$\sigma_{\Delta T} = \sqrt{<(\Delta T_j - <\Delta T>)^2>}, \quad \Delta T_j = T^{(j)}_{opt,predicted} - T^{(j)}_{opt}. \tag{2}$$

Figure 3 presents the histogram of the values of $\sigma_{\Delta T}$ in 1000 test/train splits, with average $\sigma_{\Delta T}$ being equal to 12°C for 43 organisms. This level of precision allows us to reliably discriminate between physchro-, meso-, thermo-, and hyperthermophilic organisms (2) knowing solely the amino acid composition of their proteomes. The histogram of probability to find each of the 20 amino acids in the predictor set (based on 43 organisms in 1000 training sets) is presented in Supplementary Figure S3. Each of the amino acids from the IVYWREL set can be found in the predictor sets with the probability $p>0.85$, the most frequent amino acids being Val, Tyr, and Arg..

Earlier, it has been suggested that the fraction of charged amino acid residues (Asp, Glu, Lys, Arg, DEKR) can be a significant predictor of living temperature (9, 12-15). We found that the fraction of these residues predicts the OGT with an accuracy (root-mean-square deviation) of 21°C for 86 proteomes or 14°C for 204 species. Similarly, a set of hydrophobic residues (Ile, Val, Trp, Leu, IVWL) predicts the OGT with an accuracy of 16.8°C for 86 proteomes. The IVYWREL predictor we discovered is accurate up to 8.9°C for 86 proteomes. The result clearly demonstrates that consideration of both charged and hydrophobic residues is crucially important for predicting the living temperature and, thus, thermostability.

**Membrane proteins and specific folds**

The calculation above made use of the complete proteomes, considering soluble and membrane proteins together. It is well known, however, that membrane proteins are markedly different from soluble ones, especially in terms of their stability and folding mechanism (26). Interactions with the lipid bilayer presumably result in the mechanisms of thermostabilization of membrane proteins different from those of soluble ones. We used the TMHMM hidden Markov model (22) to identify α-helical membrane proteins in the proteomes of the 83 organisms, and performed the same analysis as for the complete proteomes. The IVYWREL combination is the best predictor of thermostability of membrane proteins with three or more transmembrane helices, $R=0.89$, and is among the 5 highest-$R$ predictors for membrane proteins with ten or more helices, $R=0.85$ (see legend to Figure 4). In the latter case, the best predictor is IVYWRELGKP, $R=0.86$. Figure 4 shows that the fraction of IVYWREL amino acids in membrane alpha-helical proteins of mesophilic organisms is always higher than the overall IVYWREL fraction, whereas in hyperthermophilic organisms the fractions of IVYWREL in soluble proteins and in membrane α-helical bundles practically coincide. The weaker dependence of IVYWREL content on temperature suggests that the mechanism of thermostabilization of alpha-helical membrane proteins is different from that of soluble proteins (Note however that IVYWREL compositions for soluble and α-helical membrane proteins appear to converge in hyperthermophiles). Next, we used the ProfTMB (23) server to identify transmembrane β-barrel proteins in the proteomes of 29 organisms uniformly covering the temperature scale (see Supplementary Table S3). The most predictive amino acid combination for thermostability of transmembrane beta-barrels is CVYP, $R$=0.72, while IVYWREL is a very poor predictor, $R$=0.35. The low slope of IVYWREL correlation with OGT in α-membrane proteins and the CVYP predictor in TM β-barrels suggest that thermal adaptation in membrane proteins is governed by different rules than in globular ones (and probably different between different types of membrane proteins).

Table 2 shows that thermostability predictors of most of the folds are very similar to universal combination IVYWREL. The most abundant protein folds, such as α/β- and β-Barrel, Rossman fold, and bundle, reveal a high correlation coefficient between IVYWREL content in their sequences and OGT. Importantly, protein folds that invoke complementary mechanisms of stability, involving heme or metal-binding (globin, cytochrome C, ferredoxin) or S-S bridges (lysozyme), show a significantly lower correlation between IVYWREL content and OGT.

**IVYWREL is not a consequence of nucleotide composition bias**

Proteins are encoded in the nucleotide sequences of their genes, and thermal adaptation presumably leads to increased stability of both proteins and DNA. Therefore, signatures of thermal adaptation in protein sequences can be due to the specific biases in nucleotide sequences and vice versa. In other words, one has to explore whether a specific composition of nucleotide (amino acid) sequences shapes the content of amino acid (nucleotide) ones, or thermal adaptation of proteins and DNA (at the level of sequence compositions) are independent processes.

In order to resolve this crucial issue we applied the following logic. If amino acid biases are a consequence of just nucleotide biases and not protein adaptation then proteomes translated from randomly reshuffled genomes will feature similar ''thermal adaptation'' trends in amino acid composition as observed in real proteomes. . In contrast, if amino acid compositions are selected independently then such control calculation will result in apparently different amino acid ''trends'' in randomly reshuffled genomes than observed in reality. We carried out this computational experiment by randomly reshuffling nucleotides in the coding sequences of 83 genomes and translating the reshuffled genomes into protein sequences. Then, we calculated the amino acid compositions of the resulting control proteomes, and found (using the same exhaustive enumeration of all amino acid combinations) the amino acid combinations that are best correlated with temperature. In Figure 5 we present the probability to find an amino acid in 1000 highest-$R$ combinations for real proteomes and control proteomes created from reshuffled DNA. The striking difference between the two plots suggests that the increase of IVYWREL with environmental temperature is not a consequence of the nucleotide composition bias of coding DNA. This result proves the existence of a specific adaptation pressure acting on the amino acid composition of proteins in the process of thermal adaptation.

**Thermal adaptation of DNA**

In Figure 6a, we plotted the fraction of the proteomic thermostability predictor, IVYWREL, against the G+C content, a major contributor to pairing interactions in DNA, and a presumable indicator of DNA thermostability. A very weak negative correlation ($R$=-0.14) between the fractions of IVYWREL in the proteome and that of G+C in the coding DNA suggests that protein thermostability and its IVYWREL predictor are not consequences of enhanced GC content.

We used a complete enumeration of the $2^4-2=14$ possible sets of nucleotides to look for possible composition-OGT relationships in the coding DNA of prokaryotes, Supplementary Table S4 (on average, coding DNA constitutes 85% of prokaryotic genomes). We found that the fraction of G+C in the complete genomes of 83 (Figure 6b) or 204 (Supplementary Figure S4) species does not show any significant correlation with optimal growth temperature. The same holds true for complete genome sequences, including noncoding DNA (data not shown).

The only combination of nucleotides whose fraction is statistically significantly correlated with temperature is purine composition, A+G, $R$=0.60 (Figure 7a), in agreement with earlier observations made for several genomes (19). As this correlation is very significant, there is a possibility that the trends in amino acid and nucleotide composition are tightly linked: either the proteomic predictor, IVYWREL, is a direct consequence of the A+G nucleotide bias (the possibility that we already discarded, see above), or vice versa, the A+G nucleotide bias automatically follows from the prevalence of IVYWREL in the proteomes of thermophilic organisms.

**Purine loading bias is mainly due to IVYWREL**.

To what extent can the increase of A+G with temperature be explained by the trends in amino acid composition? Obviously, the only way to adjust nucleotide sequences without affecting the encoded proteins is through the use of codon bias. A particularly important question is whether a specific codon bias is required to reproduce the trends in DNA composition.

To answer this question, we reverse-translated the protein sequences of the 83 organisms into DNA sequences without codon bias, i.e. by using synonymous codons with equal probabilities. As shown in Figure 7b, the fraction of A+G in the resulting nucleotide sequences is very significantly correlated with the optimal growth temperature, $R=0.48$, which is very close to the corresponding value in actual genomes, $R=0.60$ (Figure 7a). In other words, the amount of variation of A+G explained by the peculiarities of amino acid composition is almost the same with and without codon bias. (We note however that the slopes of dependencies in Figs 7a,b are somewhat different suggesting that codon bias may be partly responsible for the overall purine composition of DNA). Therefore, we conclude that the fraction of A+G in the coding DNA is largely defined by the composition of proteins. Indeed, by imposing the correct amino acid composition and choosing the available synonymous codons with equal probabilities, one arrives at the correct prediction of the trend in DNA composition, increase of A+G with temperature. Together with the apparent irrelevance of G+C content for thermal adaptation on the organism scale, this finding suggests that on the level of nucleotide composition, direct selection pressure on DNA sequence composition appears to be weak.

**Nearest-neighbor correlation in DNA sequences**

After nucleotide composition, the next level of description of DNA sequences is the pairwise nearest-neighbor correlation function, or normalized probability to find successive pairs of specific nucleotides. Although sequence correlations have a minor effect on base pairing, they determine the strength of the stacking interactions, where successive A and G nucleotides (ApG pairs) have a low energy, stabilizing the DNA (27, 28).

For each of the 83 genomes, we calculated the correlation function $c_{ij}$ (see Methods) of all 16 possible successive pairs of nucleotides in the coding strand. For each pair, we plotted its $c_{ij}$ value in the 83 genomes against the OGT, and calculated the correlation coefficient, see Table 3, Column 1. It turns out that the correlation function $c_{AG}$, or the excess probability to find successive pairs of A and G nucleotides (ApG pairs) in the coding DNA is significantly increasing with OGT, $R=0.68$. The increase of $c_{AG}$ with temperature is observed for both coding DNA and the complete genome sequence.

We have shown above that codon bias does not define the dominant trends in nucleotide composition of coding DNA. Is codon bias necessary to explain the observed sequence correlations in coding parts of DNA, given that amino acid bias is established independently? To answer this question, we reverse-translated the actual protein sequences into nucleotide sequences, using synonymous codons with equal probabilities.

In principle, there are two possible sources of the correlations in DNA sequences, one stemming from the neighboring nucleotides within a codon, and another one stemming from the combination of nucleotides at the interface of successive codons. The latter possibility implies the dependence of correlations in DNA on the sequence correlations in proteins. Reshuffling of protein sequences while retaining the actual codons used for each amino acid removes the effect of codon interface (Table 3, Column 2). These data show that ApG pairs are still highly correlated with OGT, and correlations in protein sequences have a small effect on the correlations in DNA.

After removal of codon bias, ApG pairs are no longer correlated with OGT, $R=0.20$ see Table 3, Column 3 and Figure 8. Therefore, the correlation between the ApG pairs and temperature is entirely due to the evolved codon bias. Importantly, C/T pairs in coding sequences of DNA are also highly correlated with OGT. As we analyze correlations in the sense strand of DNA molecule, abundance of C/T pairs points out to the enrichment of anti-sense strand with ApG ones. Therefore, double-stranded DNA is stabilized by stacking interactions provided by ApG pairs that are spread in different locations of both sense and anti-sense strands. This conclusion holds for the whole DNA when both coding and non-coding parts are considered. Therefore, it appears that the crucial role of codon bias is to increase the number of ApG pairs in coding DNA in response to elevated environmental temperatures, enhancing the stacking interactions in DNA. We also note that the trend to increase the frequency of ApG pairs via codon adaptation may be another factor (besides amino acid adaptation, see above) that gives rise to increased overall composition of purine nucleotides A+G in hyperthermophiles, as can be seen from comparison of slopes of scatter plots shown in Figs.7a,b.

**Discussion**

Earlier works (29-31) have established an empirical correlation between OGT of an organism and the melting temperature of its proteins. Here we found the pronounced amino acid biases and related proteomic determinant of thermal adaptation: IVYWREL combination of amino acid residues, which are highly correlated with the OGT of prokaryotes. In order to better understand possible molecular mechanisms responsible for the observed highly significant proteomic trends, one must establish a connection between thermal adaptation at organismal and molecular levels.

Remarkably, all aminoacids from the IVYWREL predictor set are attached to tRNA by class I aminoacyl-tRNA synthetases (32). Other amino acids belonging to the group of class I synthetases are Cys (prone to form stabilizing S-S bridges), Lys (important in the generic mechanism of thermostability, see below and (33)), and Met that are often placed in N-termini of proteins. Thus, class I amino acids, contrary to those of class II (34), may constitute a group of amino acids sufficient for the synthesis of thermostable proteins. Besides, all class I synthetases have the Rossmann fold structure, one of ancient LUCA domains (35) with high compactness and, therefore, stability (36). These observations suggest a possible connection between thermal adaptation and evolution of protein biosynthesis.

As it is known from statistical mechanics and the theory of protein folding (37), the stability of proteins is largely determined by the Boltzmann weight $\exp(-\Delta E/k_B T)$, where $\Delta E$ is the energy gap between the native state and the nearest in energy misfolded decoy structure, $T$ is environmental temperature, and $k_B$ is the Boltzmann constant. As IVYWREL amino acids are important for thermostability, it is natural to assume that the energy gap $\Delta E$ is established mainly by interactions between these types of residues. Thus, we suggest that natural selection adjusted the content of IVYWREL in the proteomes to maintain $\Delta E/k_B T$ at a nearly constant level irrespective of the environmental temperature. Interestingly, the total fraction of IVYWREL residues in the proteomes changes from 0.37 to 0.48 over the accessible temperature range. This relatively small yet significant change highlights a very delicate balance between hydrophobic, van der Waals, ionic, and hydrogen-bond interactions in correctly folded proteins. Indeed, the IVYWREL set contains residues of all major types, aliphatic and aromatic hydrophobic (Ile, Val, Trp, Leu), polar (Tyr), and charged (Arg, Glu), both basic and acidic. Recently, we have shown, using exact statistical mechanical models of protein stability, that the increase of the content of hydrophobic and charged amino acids can be quantitatively explained as a physical response to the requirement of enhanced thermostability ("from both ends of the hydrophobicity scale" mechanism (38)), reflecting the positive and negative components of protein design.

The basic mechanism of protein thermal stability can be further augmented or modified by other biological, evolutionary, and environmental inputs. Availability of different amino acids in the environment, protein-protein interactions, and protein function requirements are just a few of those other major factors affecting the amino acid composition. Membrane proteins deliver an example where the mechanism of thermal adaptation is adjusted according to the specificity of the structure and interactions with its hydrophobic environment. In α-helical bundles, the temperature dependence of IVYWREL deviates from that of soluble proteins, whereas in β-barrels even the determinant of thermostability, CVYP, is completely different. These observations are suggestive of the differences between the folding processes of globular and membrane proteins. Thermostability predictors derived for individual folds confirm a generic nature of IVWREL predictor, which dominates in the majority of protein folds. Specific mechanisms of structure stability present in metal-binding proteins (globin, cytochrome C, ferredoxin) or those stabilized by S-S bridges (lysozyme) explain lesser importance of IVYWREL predictor for these folds.

Thermostability of proteins is, however, only a partial prerequisite of thermal adaptation of an organism, as DNA molecules must also remain stable at elevated environmental temeperatures. Numerous experimental and theoretical works established two fundamental interactions in double-stranded DNA, base pairing and base stacking as major determinants of DNA stability in vitro. GC pairs in the double helix have stronger base-pairing interactions than AT pairs (21), while purines, A and G, yield a low energy of stacking (27, 28, 39).

The role of G+C-content in establishing certain biases in the amino acid composition has been widely discussed (10, 18, 40-42). Our high-throughput analysis of nucleotide and amino acid compositions has confirmed that the G+C content of DNA is not correlated with the optimal growth temperature of prokaryotes. The only signature of thermal adaptation in DNA composition that we found, increase of A+G content in the coding DNA, is to a considerable extent a consequence of the thermal adaptation of protein sequences. Elimination of codon bias and reverse translation of the protein sequences into nucleotide sequences does not change the major trends in DNA composition (Figure 7). This result suggests that composition-dependent (pairing) interactions in double-stranded DNA are not the bottleneck of thermal adaptation in prokaryotes, as even without codon bias the variation in composition of DNA follows the variation of composition of proteins across the whole range of environmental temperatures.

On the contrary, pairwise nearest-neighbor correlations observed in DNA sequences are largely independent from amino acid trends as they are entirely determined by the codon bias developed as a form of thermal adaptation in prokaryotes. We found that natural selection tailored the codon bias to increase the fraction of ApG pairs in both strands of DNA molecules of thermophilic organisms. Indeed, an increased number of ApG pairs leads to a lower stacking energy, and, thus, stabilization of DNA. We also demonstrated that the trend of higher frequency of ApG pairs in thermophiles persists for the whole DNA, including its coding and non-coding parts. To our knowledge, this consideration is one of the first physical models that quantitatively explains the necessity and one of possible reasons for codon bias in prokaryotes.

An important finding presented in this work is that amino acid composition adaptation is a primary factor while certain signatures of thermal adaptation at the level of nucleotide composition such as purine loading index may be partly derivative of amino acid adaptation requirement. This is perhaps not surprising given that many (but not all) proteins are present in cytoplasm in monomeric form and must be stabilized on their own - by interactions between their own amino acids (and also destabilize misfolded conformations, performing negative design as well). On the other hand many proteins bind to DNA in prokaryotic cells and some of them may provide additional stabilization to DNA in hyperthermophiles making thermal adaptation at the level of DNA sequences a ''collective enterprise'' of the cell, relieving direct pressure on DNA sequence to adapt to high temperature. Positive superhelicity provided by gyrases may be one such mechanism and perhaps several other mechanisms of DNA stabilization by proteins could be discovered in future. In particular the observed purine-purine AG correlation as an ''independent'' (from amino acid composition) adaptation mechanism may be a signature of the important role of stacking interaction in DNA thermal stabilization. Alternatively, the AG correlations may be a signature of an additional thermal adaptation of DNA via modulation of local mechanical properties or facilitation of interaction with proteins. Elucidation of the exact mechanism(s) by which DNA in prokaryotic genomes adapt to high temperatures is a matter of future research.

To summarize, we have shown that thermal adaptation of proteins in prokaryotes is strongly manifest at the level of amino acid composition. Among all possible combinations of amino acids, the fraction of amino acids IVYWREL in a proteome is the most precise predictor of optimal growth temperature. We also observe a difference in the signatures of thermal adaptation of soluble and membrane proteins. We did not find a strong evidence for thermal adaptation of DNA at the level of nucleotide composition. The G+C content of the genomes is not correlated with environmental temperature, while A+G content increases with environmental temperature, to a considerable extent, as a consequence of thermal adaptation of proteins. At the same time, our analysis of sequence correlations in DNA shows that ApG nearest-neighbor pairs are overrepresented in both strands of dsDNA of thermophilic organisms. The abundance of ApG pairs is a direct consequence of the codon bias developed in thermophilic prokaryotes. These findings provide a definitive answer to the long-standing problem of which genomic and proteomic compositional trends reflect thermal adaptation.

**Acknowledgements**

We are grateful to M.D. Frank-Kamenetskii for illuminating discussions and M. Kloster for useful comments on initial version of this paper. We would like to thank an anonymous referee for the suggestion to analyze membrane proteins and to H. Bigelow (Columbia University) for the help with prediction of TM beta-barrels. This work was supported by the NIH. I.N.B. was supported by the Merck fellowship for genome-related research.

**Figure captions**

**Figure 1.**

Correlation between the sum $F$ of fractions of Ile, Val, Tyr, Trp, Arg, Glu, and Leu (IVYWREL) amino acids in 86 proteomes and the optimal growth temperature of organisms $T_{opt}$. The linear regression (red line) corresponds to the correlation coefficient $R=0.93$. The optimal growth temperature $T_{opt}$ (in degree Celsius) can be calculated from the total fraction $F$ of IVYWREL in the proteome according to $T_{opt}=937F-335$. By construction, the IVYWREL set is the most precise predictor of OGT among all possible combinations of amino acids; other combinations statistically yield a larger error of prediction of OGT.

**Figure 2**

Distribution of the correlation coefficient $R$ between OGT and fractions of amino acids in a proteome for different variation of the IVYWREL set, additions or deletions of one or two amino acids to/from the set, or substitution of one or two amino acids from the set by one or two amino acids not from the set. The dashed red line at $R=0.93$ corresponds to the unperturbed IVYWREL. The horizontal red lines indicate the median values of the correlation coefficient for the given type of change of the predictor set.

**Figure 3**

The distribution of the root-mean-square error $\sigma_{\Delta T}$ of the prediction of the OGT in the 1000 43-species test sets (black), and the prediction errors for 86 organisms using the IVYWREL set (red) or sets of charged or hydrophobic residues (blue). Individually, sets of hydrophobic or charged residues provide a much lower precision than their proper combination, IVYWREL.

**Figure 4**

Total fraction of IVYWREL amino acids in alpha-helical membrane proteins containing three or more (green) or ten or more (red) helices, plotted against the OGT. Solid lines are linear regressions. The black line is the linear regression of the fraction of IVYWREL amino acids in all proteins in a proteome, same as in Figure 1. The fraction of IVYWREL in membrane proteins of thermophilic organisms is about the same as in all of their proteins. In mesophiles, membrane proteins are enriched with hydrophobic residues. For membrane proteins with ten or more helices, the five best predictors are ILVWYGERKP, FILVWYATSERKP, FILVWYATSEHRKP, IVYWREL, MFILVWYATSEHRKP, $0.85 < R < 0.86$.

**Figure 5**

Histogram of the probability to find an amino acid among 1000 combinations of amino acids which are most correlated with OGT for real proteomes (red) and for artificial proteomes created from reshuffled DNA sequences (blue). The histogram for real proteomes supports the stability of IVYWREL predictor, while the difference between the two histograms suggests that amino acid biases upon thermophilic adaptation are not a consequence of the trends in nucleotide composition.

**Figure 6**

**(a)** Dependence of the fraction of IVYWREL amino acids in 83 proteomes (protein thermostability predictor) on the fraction of G+C in the coding DNA in the corresponding genomes.
**(b)** Dependence of the G+C content in the coding DNA of the 83 complete genomes on the optimal growth temperature of the organisms. The correlation coefficient is $R=-0.15$, indicating that G+C content of the coding DNA is not related to the optimal growth temperature.

**Figure 7**

**(a)** The fraction of A+G in the coding DNA of the 83 complete genomes is highly correlated with the optimal growth temperature, $R=0.60$.
**(b)** When protein sequences of the 83 organisms are reverse translated into DNA without codon bias, the fraction of A+G remains correlated with OGT, $R=0.48$

**Figure 8**

Dependence of the pairwise nearest-neighbor correlation function for A,G nucleotides $c_{AG}$ on the optimal growth temperature in the genomes of 83 species (black) and in the DNA sequences obtained from the proteomes of 83 species without codon bias (red). While codon bias is not essential for reproducing the trends in nucleotide composition (Figure 6), nucleotide correlations are entirely dependent on the proper choice of codon bias. An increase of the number of ApG pairs enhances the stacking interactions in DNA, stabilizing it at elevated environmental temperatures.

**Table Captions**

**Table 1.**

Effects of the changes in the IVYWREL predictor combination on its predictive power. The plus sign means that an amino acid is added to the set, the minus sign indicates removal of an amino acid, "subst" indicates substitution.

**Table 2**

Thermostability predictors for the major protein folds. Column 1, fold; column 2, number of species (out of 83) where proteins with that fold have been detected; column 3, average number of detected proteins per genome; column 4, maximum correlation coefficient between sum of fractions of amino acids and the corresponding amino acid combination; column 5, correlation coefficient between fraction of IVYWREL and OGT for the proteins with a given fold.

**Table 3**

Correlation coefficients $R$ between OGT and DNA sequence correlation function $c_{ij}$ for the 16 combinations of nucleotides $i,j$ in 83 genomes. Column 1, actual DNA sequences; column 2, DNA sequences obtained from reshuffled protein sequences retaining the actual codons used for every amino acid; column 3, DNA sequences obtained from reshuffled proteins without codon bias. ApG pairs are the most correlated with OGT in real genomes, but this property vanishes if codon bias is removed. Each of the columns is ordered by the value of $R$.

**Table 1**

| Type of change | $R_{median}$ | Worst case | | Best case | |
|---|---|---|---|---|---|
| | | $R_{min}$ | Change | $R_{max}$ | Change |
| IVYWREL+1 | 0.877 | 0.47 | +A | 0.921 | +M |
| IVYWREL+2 | 0.804 | 0.24 | +AQ | 0.917 | +FP |
| IVYWREL-1 | 0.855 | 0.64 | -I | 0.921 | -W |
| IVYWREL-2 | 0.754 | 0.40 | -IE | 0.874 | -WE |
| IVYWREL subst. 1 | 0.776 | 0.18 | E→A | 0.914 | W→H |
| IVYWREL subst. 2 | 0.580 | -0.23 | VE→AQ | 0.902 | WR→GP |

**Table 2**

| Fold | Organisms | Proteins/Org | $R_{max}$, predictor | $R_{IVYWREL}$ |
|---|---|---|---|---|
| Beta barrel | 83 | 79.2 | 0.91 IVWYAQERKP | 0.87 |
| Beta helix | 59 | 20.3 | 0.90 ILVWYGERKP | 0.81 |
| Bundle | 83 | 197 | 0.88 MILVWYANE | 0.82 |
| Globin | 78 | 7.2 | 0.78 VWYGERKP | 0.53 |
| Cytochrome C | 55 | 5.7 | 0.72 ILVWYGNQERKP | 0.44 |
| Ferredoxin | 83 | 99 | 0.83 CMFILVWYGEHRKP | 0.45 |
| Lysozyme | 50 | 4.2 | 0.72 CFILVWYGNDEP | 0.50 |
| Rossman fold | 83 | 292 | 0.90 ILVWERKP | 0.86 |
| Sandwich | 82 | 27 | 0.85 FILVWYGNERKP | 0.74 |
| TIM-barrel | 82 | 48 | 0.87 ILVWYGERKP | 0.87 |

**Table 3**

| Real DNA | Reshuffled proteins, real codon bias | No codon bias, reshuffled proteins |
|---|---|---|
| -0.589 TG | -0.597 TG | -0.825 CA |
| -0.515 CA | -0.459 AA | -0.698 TG |
| -0.458 GC | -0.458 CG | -0.563 GC |
| -0.443 AT | -0.456 GC | -0.437 AC |
| -0.432 CG | -0.452 AT | -0.355 AT |
| -0.302 AA | -0.388 CA | -0.127 TT |
| -0.202 GT | -0.244 GT | 0.082 GG |
| -0.076 AC | 0.010 TC | **0.177 AG** |
| 0.092 TC | 0.097 AC | **0.216 CT** |
| 0.167 TT | 0.187 TT | 0.275 TC |
| 0.200 GA | 0.208 GA | 0.334 GA |
| 0.392 TA | 0.446 GG | 0.343 TA |
| 0.479 GG | 0.456 TA | 0.396 AA |
| 0.558 CC | 0.567 CC | 0.417 GT |
| **0.601 CT** | **0.574 CT** | 0.443 CG |
| **0.680 AG** | **0.736 AG** | 0.868 CC |

**Figure 1**

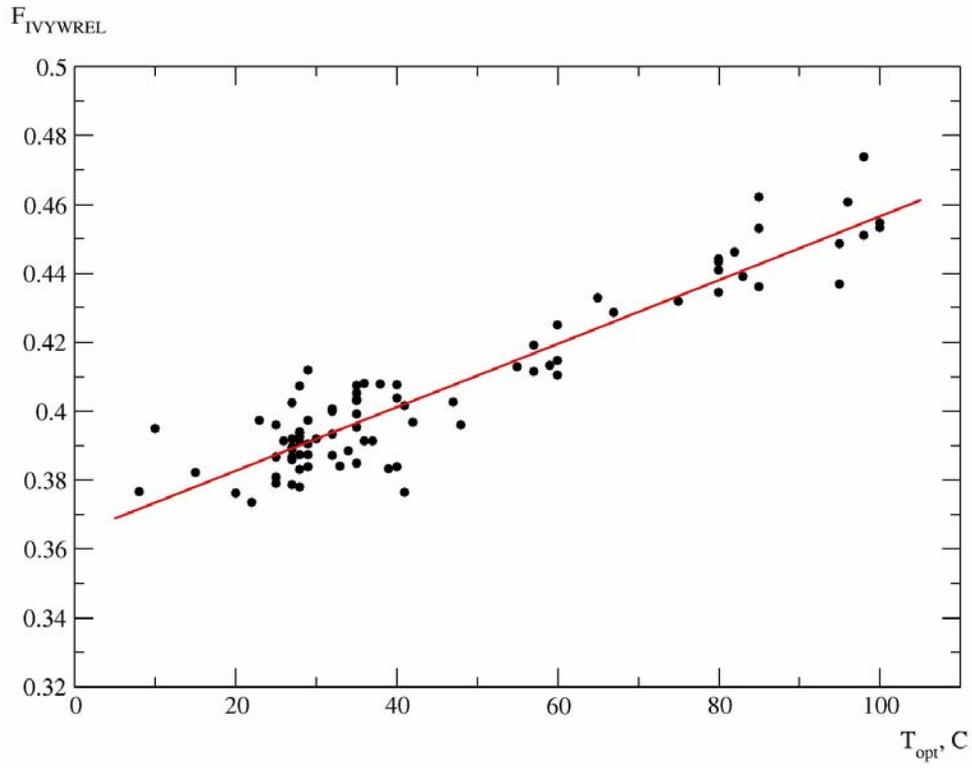

**Figure 2**

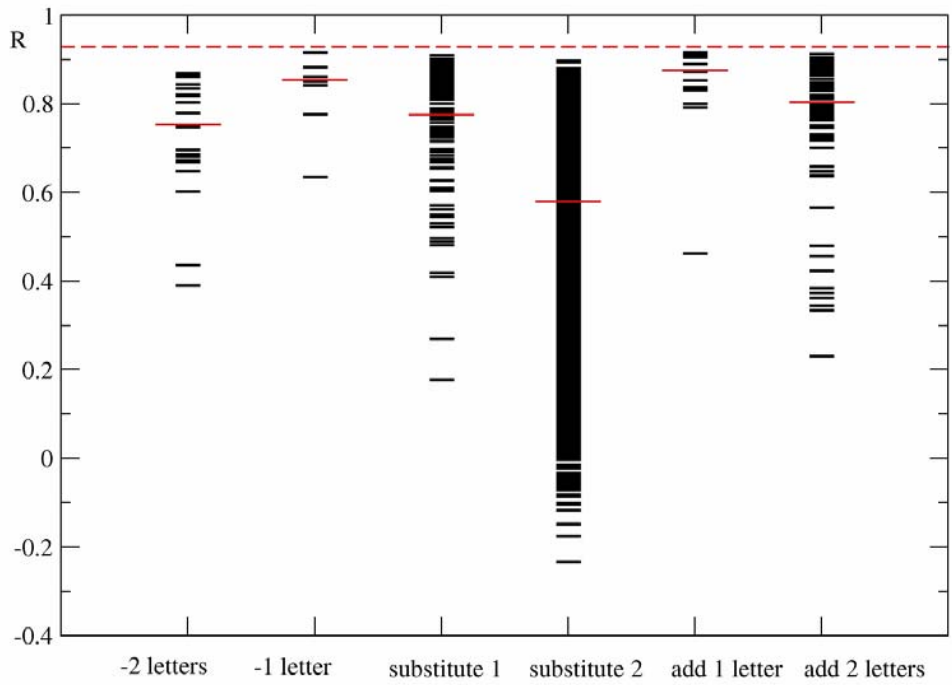

**Figure 3**

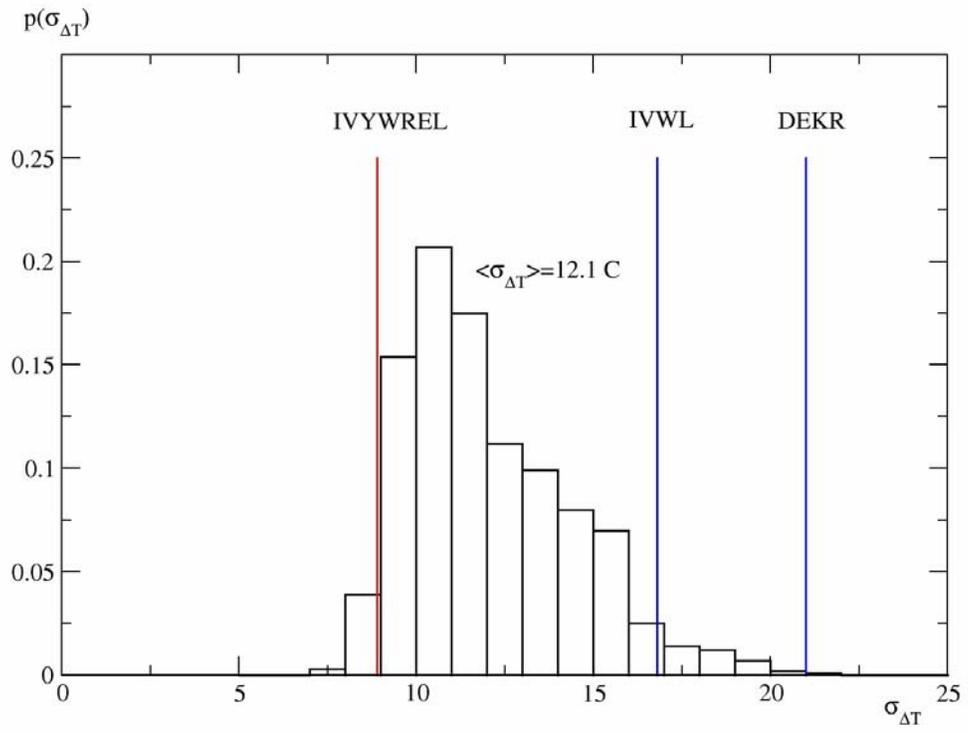

**Figure 4**

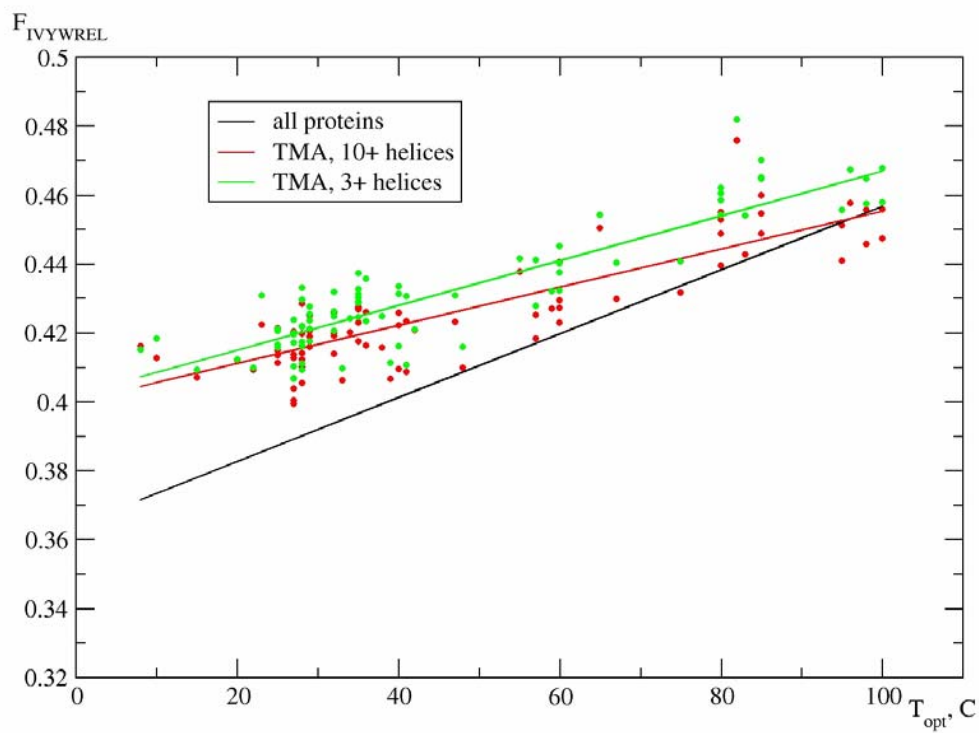

**Figure 5**

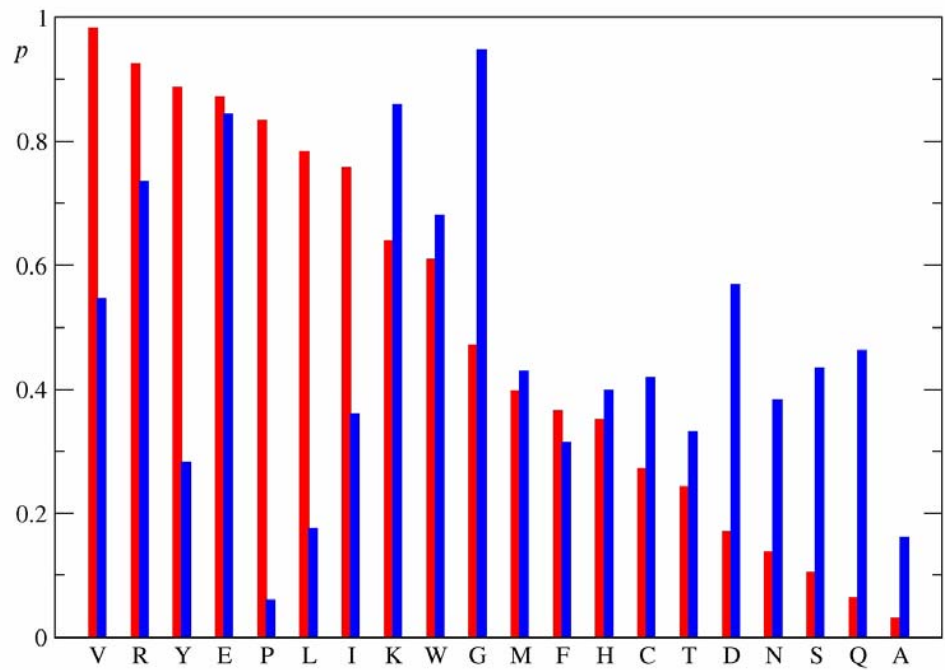

**Figure 6a**

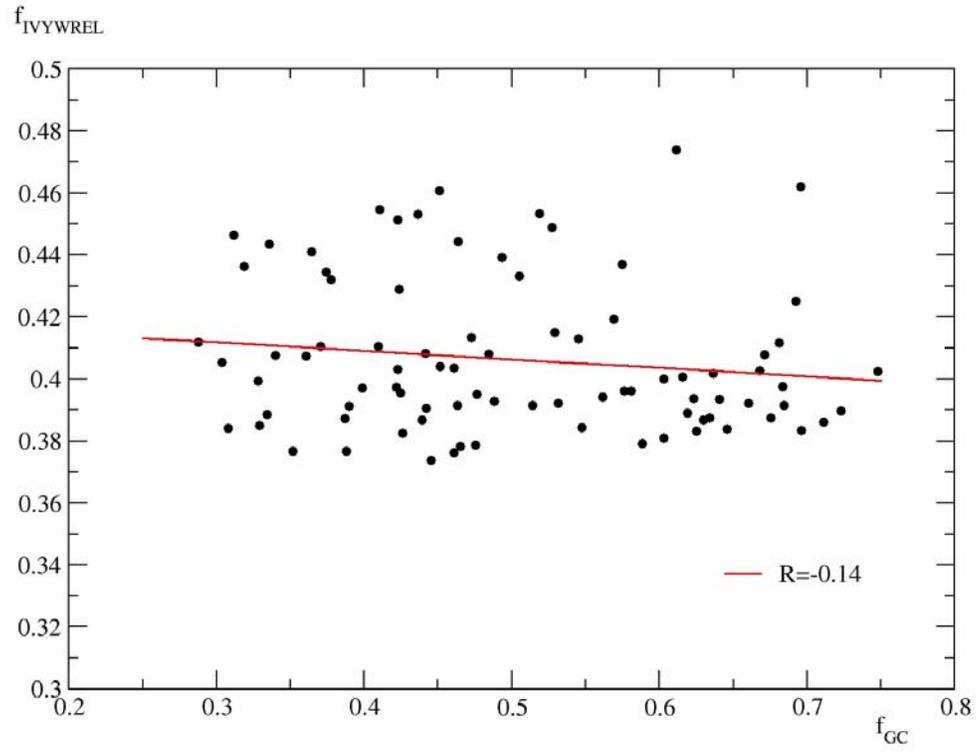

**Figure 6b**

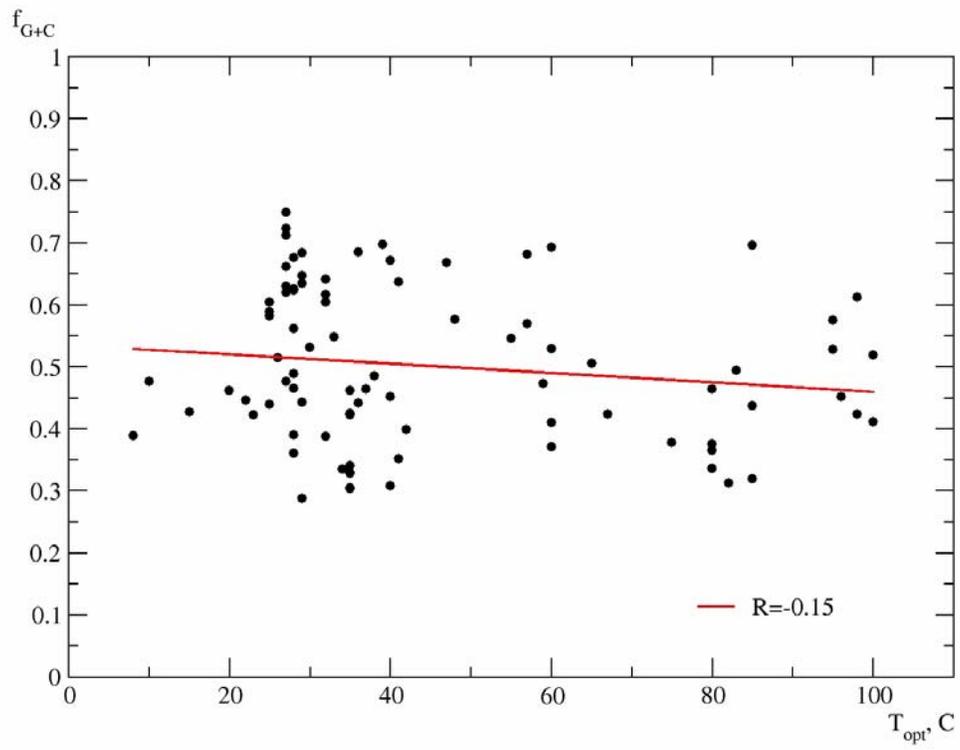

**Figure 7a**

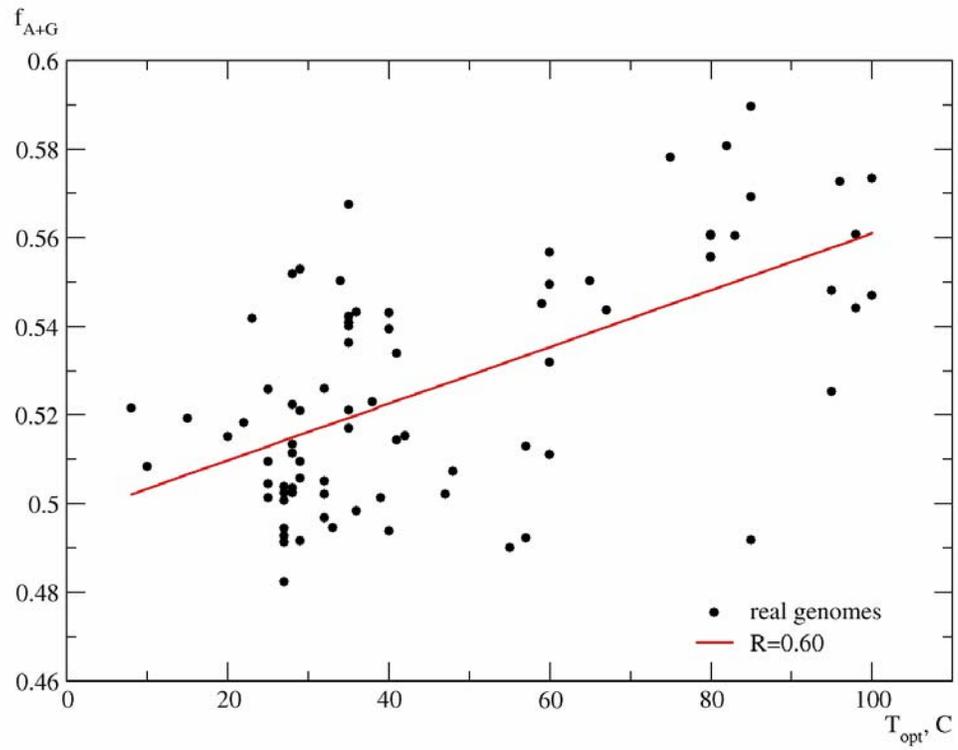

**Figure 7b**

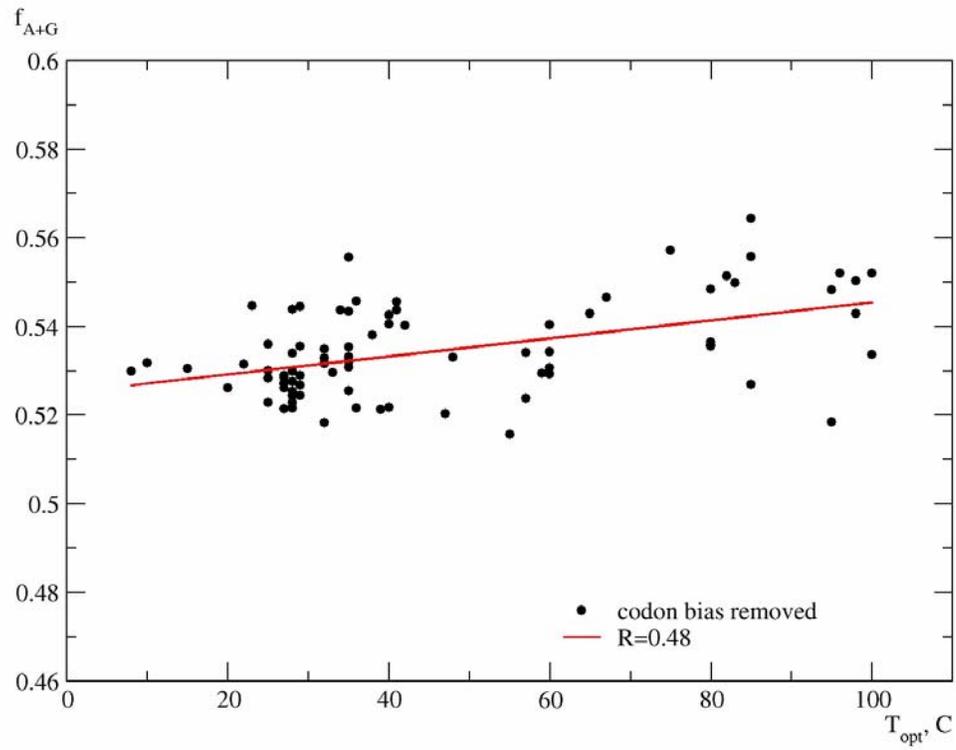

**Figure 8**

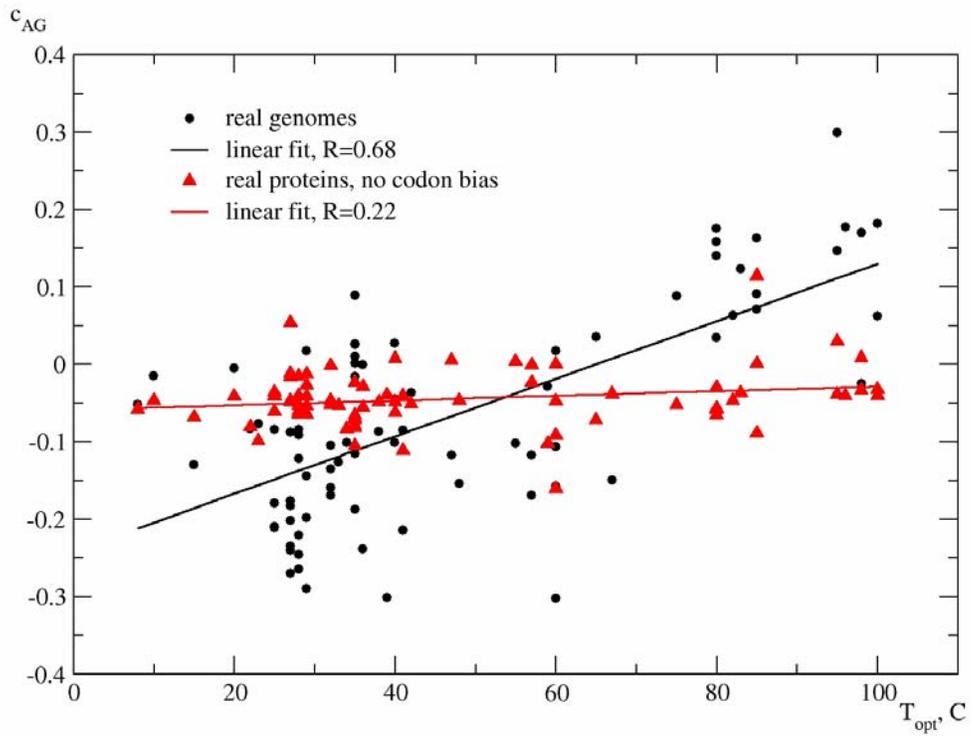